\documentclass[letterpaper,1wpt]{report}

\title
{
	A Secure Web-Based File Exchange Server\\
	Software Requirements Specification Document
}

{\author
	{\bf
		{CIISE Security Investigation Initiative}\\\hline\\
		Represented by:\\\\
		Serguei A. Mokhov\\
		Marc-Andr\'e Laverdi\`ere\\
		Ali Benssam\\
		Djamel Benredjem\\
		\texttt{\{mokhov,ma\_laver,d\_benred,al\_ben\}@ciise.concordia.ca}
		\\\\\\
		Montr\'eal, Qu\'ebec, Canada\\\\\\
	}
}

\date{December 14, 2005}

\usepackage{graphicx}
\usepackage{latexsym}
\usepackage{makeidx}
\usepackage{url}

\makeindex

\newcommand{\lucidL}[1]{{$\mathit{Lucid}$}($L$) }

		{}

\def\myvert{\raise 2.27pt \hbox{\vrule depth 0pt height 8pt width 0.2mm}}
\def\myarrow{\hspace*{0.43mm}%
             \raise 2.29pt\hbox{\vrule depth 0pt height 8pt width 0.16mm}%
             \hspace*{-0.32mm}%
             $\longrightarrow$
             \ %
             }

\begin{document}

	\begin{titlepage}
		\maketitle
	\end{titlepage}

	\pagenumbering{roman}
\tableofcontents
\clearpage
\pagenumbering{arabic}

\listoffigures

	\chapter{Introduction}
\index{Introduction}



Building Trust is the basis of all communication, especially
electronic one, as the identity of the other entity remains
concealed. To address problems of trust, authentication and security
over the network, electronic communications and transactions need a
framework that provides security policies, encryption mechanisms and
procedures to generate manage and store keys and certificates.

This software requirements specification (SRS) document demonstrates
all the concerns and specifications of the secure web-based file
exchange server (SFS). SFS is a security architecture that we
propose here to provide an increased level of confidence for
exchanging information over increasingly insecure networks, such as
the Internet. SFS is expected to offer users a secure and
trustworthy electronic transaction.

\section {Purpose}
The intent of implementation and deployment of  SFS facilities is to
meet its basic purpose of providing Trust. Presently, SFS needs to
perform the following security functions:

\begin{itemize}

\item  \emph{Mutual authentication of entities taking part in the
communication:} Only authenticated principals can access files to
which they have privileges.

\item \emph{Ensure data integrity:} By issuing digital certificates
which guarantee the integrity of the public key. Only the public key
for a certificate that has been authenticated by a certifying
authority should work with the private key possessed by an entity.
This eliminates impersonation and key modification.

\item \emph{Enforce security:} Communications are more secure by using
SSL to exchange information over the network.

\end{itemize}

\section {Scope}
SFS is implemented to secure sensitive resources of the organization
and avoid security breaches. The SFS allows trustworthy
communication between the different principals. These principals
must be authenticated and the access to the resources (files) should
be secured and regulated. Any principal wants to access to the
database needs to perform the following steps:

\begin{itemize}

\item  \emph{Mutual authentication:} The Web Server via which the database is contacted authenticates the
principal using its digital certificate and username to ensure that
it is who it claims to be . The principal authenticates also the
server using its certificate information.

\item  \emph{Principal validation:} To validate the principal, the
server looks up information from an LDAP server which contains the
hierarchy of all principals along with certificates and credentials.

\item  \emph{Enforcing security:} The security is enforced by using SSL to communicate between the
Web Server and the LDAP server, the Web Server and the database and
between the principal and Web Server.

\item  Principal authentication: Upon successful authentication,
the Web Server will allow the principal to perform actions on the
database according to a pre specified Access Control List.

\item  Kinds of principals: There are two kinds of principals, administrators and
clients: \emph{Clients} have the ability to upload, download, delete
and view files. \emph{Administrators} have the ability to: Upload,
download, delete and view files; Add, delete and modify users;
Generate user's certificate, with all required information; Generate
ACL to users; Manage groups, Perform maintenance.

\end{itemize}

Finally, this infrastructure allows additional features such as the
ability to assign users to groups in order to provide users with the
access to files prepared by other group members.

\section {Definitions and Acronyms}

\begin{itemize}

\item     PKI: Public Key Infrastructure
\item     OpenLDAP : is a free, open source implementation of the Lightweight
          Directory Access Protocol (LDAP).
\item     OpenSSL: an open source SSL library and certificate authority
\item     Apache Tomcat: A Java based Web Application container that was created to
          run Servlets and JavaServer Pages (JSP) in Web applications
\item     PostgreSQL: An open source object-relational database server
\item     SSL: Secure Socket Layer
\item     JSP: Java Server Pages
\item     JCE: Java Cryptography Extension
\item     API: Application Programming Interface
\item     JDBC: Java Database Connectivity
\item     JNDI: Java Naming and Directory Interface
\item     LDAP: Lightweight Directory Access Protocol
\item     X.509: A standard for defining a Digital Certificate used by SSL
\item     SRS: Specification Request Document
\item     SDD: Specification Design Document
\item     DER: Distinguished Encoding Rules
\item     Mutual Authentication: The process of two principals proving their identities to each other
\item     SFS: Secure File Exchange Server, this product
\item     COTS: Commercial Off The Shelf, common commercially or freely available software

\end{itemize}

The coming sections of the SRS are a description of all the
requirements to be implemented in SFS system. The requirements
specifications are organized in two major sections: Overall
Description and Specific Requirements.


%

\chapter{Overall Description}

In this chapter we provide an overall insight of the general factors
that affect the SFS system and its requirements.

\section{Product Perspective}

The SFS system is intended to operate in a distributed environment:
clients machines, application server, database server, and LDAP
server. SFS is accessed via secure connections we intend to provide
in this work. The system's user can be either a normal user or an
administrator. A normal user has the ability to upload, download,
delete and view files. An administrator is able to: Upload,
download, delete and view files; add, delete and modify users;
generate user certificates along with all required information;
generate an ACL to each and every user; manage groups; perform
various maintenance actions such as: check log files, delete files,
etc.

\subsection{System interfaces}

The various parts of the SFS application will be installed on client
machines and different servers. The client that uses the system has
to have the certificate installed on his machine to provide client
authentication. The servers are responsible of one of the following
functions: provide the database, provide the LDAP server
functionalities, and provide the application server for different
clients.

\subsection{User Interfaces}

The user interfaces consist of web-based graphical components that
allow the user to interact with the SFS system. The user will use a
web browser to send and receive data. If the user is the
administrator, s/he will have more options to add, delete, etc.
users, generate users certificates, generate ACL for each user, etc.

\subsection{Hardware Interfaces}

The hardware interfaces will be achieved through the abstraction
layer of the Java Virtual Machine (JVM). The keyboard and the mouse
are examples of such hardware interfaces that allow users to
interact with the SFS system.

\subsection{Software Interfaces}

Among the most important software interfaces used in this project,
we have:
\begin{itemize}

\item The SFS  system is OS-independent due to the cross platform Java
implementation. It will support web browsers such as Internet
Explorer, Mozilla Fireworks, etc.

\item Access to databases will be provided by JDBC 3 on both Windows
and Linux environment.

\item JXplorer? will be used to provide a graphical access to LDAP server.

\item OpenLDAP? is used to host users' certificates.

\item Java 1.5 JDK from Sun.

\item JRE 1.5 from Sun.

\item Servlets for client and administrator interfaces.

\item Apache Tomcat5 server as the web server used in this project.

\item PostegreSQL? is the database used to host users information, files,
etc.

\item OpenSSL toolkit to generate the certificates for users.

\end{itemize}

Hereafter, we provide the software and documentation's locations
related to these interfaces:

\begin{itemize}

\item OpenLDAP? software and documentation found at: \emph{http://www.openldap.org/}

\item PostgreSQL database and documentation found at: \emph{http://www.postgresql.org/}

\item  Java development kit 1.5 available at: \emph{http://java.sun.com/j2se/}

\item JSP documentation found at: \emph{http://java.sun.com/products/jsp/}

\item OpenSSL  Toolkit found at: \emph{http://www.openssl.org/}

\item Apache Tomcat 5.0 web server found at: \emph{http://tomcat.apache.org/}

\end{itemize}

\section{Product Functions}

The SFS system will implement the following functionalities:

\begin{itemize}

\item \emph{Server authentication:} This use case allow
the user to authenticate the web server is connecting to.

\item \emph{Client authentication:} This is use case allow the server
to authenticate the user he is trying to connect to.

\item \emph{Secure communication:} Between users over the network.

\item \emph{Files handling:} Such as downloading, uploading, and
deletion files.

\item Users management: Administrator has the ability to add and delete users, add and delete groups, and assign users to groups.

\end{itemize}

\chapter{Specific Requirements}

In this section we describe the software requirements to design the
SFS system. The system design should satisfy the following
requirements.

\section{Functional Requirements} Hereafter, we express the
expectations in terms of system functions and constraints. This
includes the domain model and the most important use case diagrams
of the SFS system.

\subsection{Domain Model}

The SFS system domain model consists of many packages.

\begin{itemize}

\item \emph{Client authentication module:} used to provide users the
ability to authenticate the server.

\item \emph{Server authentication module:} used to provide the server the
ability to authenticate the clients.

\item \emph{LDAP connection module:} used to provide connection to the
LDAP server in order to check clients' credentials.

\item \emph{Database connection module:} used to provide users the
ability to connect to the database server in a secure mode.

\end{itemize}

\begin{figure}[htbp]
\begin{center}
  \fbox{
      \scalebox{0.5}{  \includegraphics{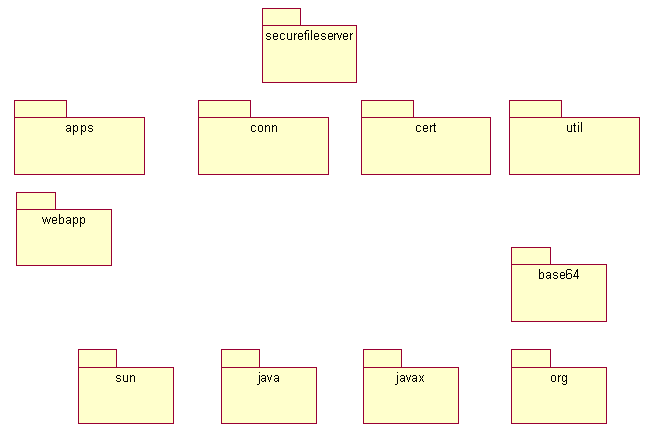}
      }
    }
    \caption{\label{fig1} SFS system packages}
\end{center}
\end{figure}

\subsection{Use Case Model}

The SFS system consists of a set of use cases manageable by the
users of the application. There are two types of users of the
system: normal users and administrator. Normal can view, delete,
download, and upload files. For the administrator, s/he can: add,
delete, etc. users; generate users' certificates; generate ACL for
each user, etc.

\begin{figure}[htbp]
\begin{center}
  \fbox{
      \scalebox{0.4}{  \includegraphics{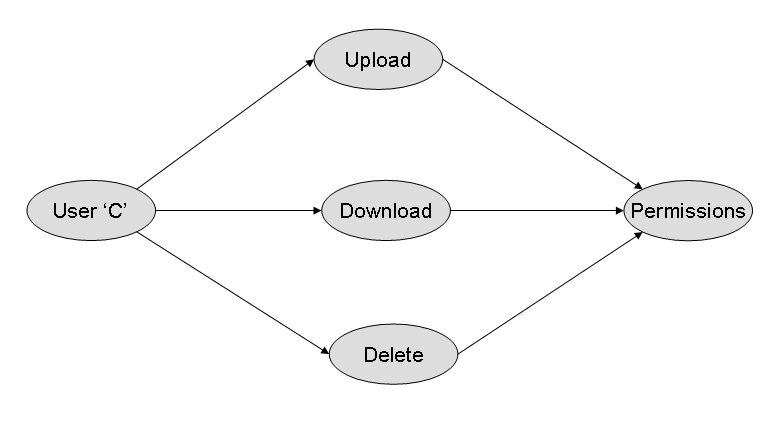}
      }
    }
    \caption{\label{useuse} Normal user use-case}
\end{center}
\end{figure}

The diagram in Figure \ref{useuse} shows the capabilities of a
normal user.

\begin{figure}[htbp]
\begin{center}
  \fbox{
      \scalebox{0.45}{  \includegraphics{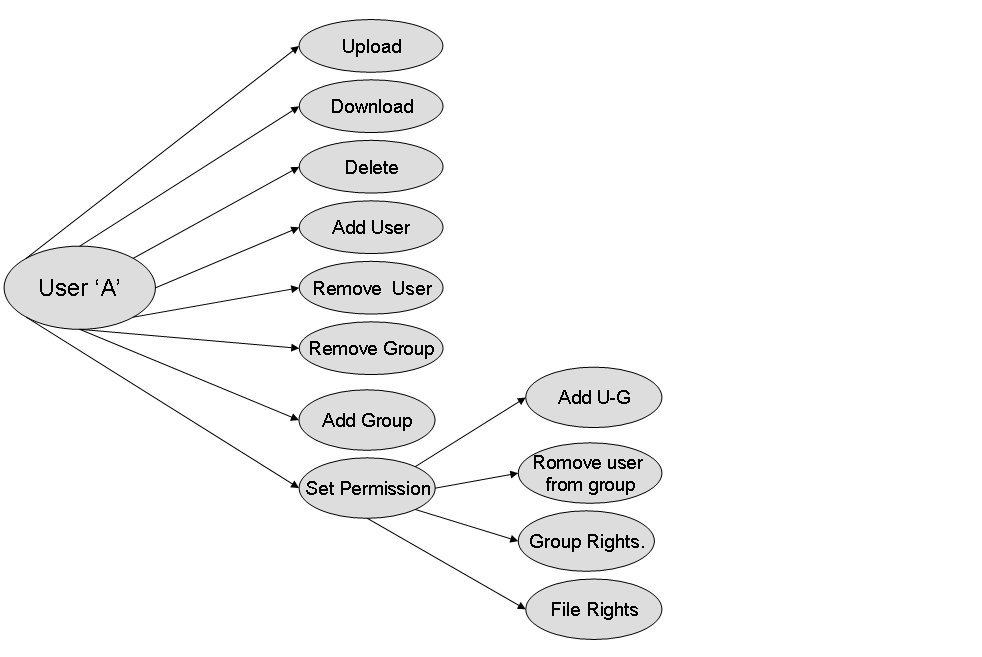}
      }
    }
    \caption{\label{servuse} Administrator use-case}
\end{center}
\end{figure}

\newpage
The diagram in Figure \ref{servuse} shows the capabilities of the
administrator user.

\section{Software System Attributes}

There are a number of software attributes that can serve as
requirements.  It is important that required attributes by specified
so that their achievement can be objectively verified. The following
items are some of the most important ones: security, reliability,
availability, maintainability, and portability.

\subsection{Security}

Security is the most important attribute of the SFS design and
implementation. The mutual authentication between the server and
clients is crucial for the system use. The system should be able to
authenticate users and differentiate among them, either are normal
users or administrator. In order to achieve the security feature
expected from the SFS system, the following tasks have to be
realized:

\begin{itemize}

\item Utilize cryptographic techniques

\item Check users' credentials before using the system and accessing the database

\item Provide secure communications between different parts of the
system

\end{itemize}

\subsection{Reliability}

The basis for the definition of reliability is the probability that
a system will fail during a given period. The reliability of the
whole system depends on the reliability of its components and on the
reliability of the communication between its components. The SFS
system is based mainly on some standard components such as OpenLDAP,
OpenSSL, JDBC, PostgreSQL, etc. The reliability of these service
components is already proved. This fact improves the reliability of
the system and restricts the proof work on assuring only of the
reliability of the added components and the communications between
the different components. In addition, the system must ensure the
security of the communications which is the most important issue of
the SFS system.

\subsection{Availability}

The SFS system must be able to work continuously in order to provide
users with an access to different server's parts of the system.
However, since this system depends on distributed information
systems and databases, many constraints should be taken into account
such as:

\begin{itemize}

\item The connection to the web server that provides access to the
system

\item The interconnection between different parts of the system should
always be available; otherwise, the users cannot complete their
tasks using the system.

\item The database should be available in the database server side

\item The LDAP server should be always available in order to check
users' credentials

\item The web server should be also available in order to allow users
connecting to the system.

\end{itemize}

\subsection{Maintainability}

Maintainability is defined as the capacity to undergo repairs and
modifications. The main goal in designing SFS system is to keep it
easy to be modified and extended.

\subsection{Portability}

The portability is one of the main specifications of Java. Since SFS
is implemented using the Java programming language, the portability
is automatically satisfied and the system is able to run on any
machine or operating system which supports the execution of a Java
virtual machine.

\section{Logical Database Requirements}

The rationale behind SFS system is to provide secure connections for
users accessing databases to view, delete, upload, and download
files through a web server and LDAP server. After analyzing the
requirements we propose using a relational database model to meet
our requirements. This database is required to store information
about users, files, groups of users, etc.  The database is expected
to work on 24 hours and 7 days in order to provide nonstop access to
the users. Therefore, backup of the database should be taken
periodically (daily or weekly. The relational database itself
guarantees the flexibility, simplicity and elimination of redundancy
once designed carefully. The entity relationship model will be
elaborated in detail in the database design of the design part, in
this document.


	\addcontentsline{toc}{chapter}{Bibliography}


\nocite{frequently-used-ssl-commands}
\nocite{tomcat}
\nocite{soen-uml-and-patterns-2006}
\nocite{eclipse}
\nocite{jxplorer}
\nocite{junit}
\nocite{oreilly-servlet-cos}
\nocite{jsp}
\nocite{debbabi-inse6120-2005}


\nocite{openssl}
\nocite{postgres}
\nocite{servlets}
\nocite{suranjan03}
\nocite{openldap}

\bibliography{common/report}
\bibliographystyle{alpha}



	\printindex
\end{document}